# Valley-selective Klein tunneling through superlattice barrier in graphene

Xing-Tao An[1*], Wang Yao[2*]

[1]School of Science, Hebei University of Science and Technology, Shijiazhuang, Hebei 050018, China
[2]Department of Physics, University of Hong Kong, Hong Kong, China

*Correspondence to: anxt2005@163.com, wangyao@hku.hk

## Abstract

Graphene electrons feature a pair of massless Dirac cones of opposite pseudospin chirality at two valleys. Klein tunneling refers to the intriguing capability of these chiral electrons to penetrate through high and wide potential barrier. The two valleys have been treated independently in the literature, where time reversal symmetry dictates that neither the normal incidence transmission nor the angle-averaged one can have any valley polarization. Here we show that, when intervalley scattering by barrier is accounted, graphene electrons normally incident at a superlattice barrier can experience a fully valley-selective Klein tunneling, i.e. perfect transmission in one valley, and perfect reflection in the other. Intervalley backscattering creates staggered pseudospin gaps in the superlattice barrier, which, combined with the valley contrast in pseudospin chirality, determines the valley polarity of Klein tunneling. The angle averaged transmission can have a net valley polarization of 20% for a 5-period barrier, and exceed 75% for a 20-period barrier. Our finding points to an unexpected opportunity to realize valley functionalities in graphene electronics.

The chiral nature of relativistic electrons results in their counterintuitive scattering behaviors at potential barriers, known as the Klein paradox. [1] The chiral

Dirac quasiparticles in graphene make possible a condensed matter testbed for this exotic consequence of quantum electrodynamics. [2] Upon normal incidence at a barrier, the Klein paradox manifests either as the perfect transmission in single-layer graphene, or the perfect reflection in bilayer graphene. This difference between the two graphene systems arises from their distinct chirality structures associated with the *sublattice pseudospin*, which dictate either forward-propagation inside the barrier or backscattering is allowed under pseudospin conservation. The electron mean free path can reach tens of microns in high quality grapheme, [3, 4] promising the exploitation of the coherent chiral tunneling effect for novel electron devices with optics analogs. [5-11]

Graphene electron also features a valley degree of freedom, labeling the two massless Dirac cones with *opposite* pseudospin chirality at the K and -K corners of Brillouin zone. The possibility to address and exploit valley as information carrier has led to the conceptual electronic applications known as valleytronics. [12-14] Valley selective flow of carriers, or valley current, is the main element to enable valley functionalities. A variety of schemes have been explored for producing valley current in graphene by introducing edges, [13] inversion symmetry breaking, [14-18] line defects and topological interfaces, [19-24] or exploiting strains [25, 26] and trigonal warping. [27] In the context of Klein tunneling, possibilities to engineer valley polarized transmission at selected oblique angles are discovered, [9, 28] but the angle-averaged transmission cannot carry valley polarization as a consequence of time-reversal symmetry, [29] limiting such schemes to ballistic devices and angle resolved operations. All the above mechanisms exploit intravalley process only, whereas intervalley scattering is generally considered as a deleterious cause of error for the valley functionalities. On the other hand, a counterintuitive role of intervalley scattering in pumping valley polarization has been revealed in recent theoretical studies. [29, 30]

Here we discover that, at a superlattice barrier in single-layer graphene, intervalley scattering can selectively block the Klein tunneling in a chosen valley while retaining the perfect transmission in the other. This valley selectivity is made possible by the staggered pseudospin gaps created by intervalley backscattering in the superlattice barrier. For electrostatically defined superlattice, the gate control of the barrier height can be used to switch the valley polarity of the Klein tunneling. The angle-averaged transmission can have a net valley polarization of 20% for a 5-period barrier, and exceed 75% for a 20-period barrier, where the total width of superlattice can be practically restricted in submicron regime using O(10) nanometer periodicity. This chiral tunneling phenomenon makes possible a high throughput valley filter, where the sizeable angle-integrated valley polarity is crucial for harvesting the filtered valley current beyond the ballistic limit.

**Staggered pseudospin gaps in graphene superlattice.** Let us consider a single-layer graphene subject to a Kronig-Penney type electrostatic potential along the zigzag direction, with square barriers of width $W$ and height $U$ arranged with periodicity $L$ (Fig. 1a). The effect of such superlattice on the massless chiral electrons has been investigated in the absence of intervalley scattering.[7] Intravalley scattering by the superlattice potential leads to anisotropic massless Dirac dispersion: the group velocity is renormalized in the armchair (*y*) direction parallel to the barriers, but unchanged in the perpendicular zigzag (*x*) direction, as a manifestation of the Klein paradox. The step-shaped potential can also introduce intervalley scattering that couples the two valleys, which, however, has not been considered in the literature.

To understand the role of intervalley scattering in such graphene superlattices, we first analyze the propagation of electron wavefunction in a single period, as Fig. 1b illustrates. For normal incidence at the barrier, intravalley backscattering is forbidden by the orthogonality in pseudospin of the initial and final states. Nevertheless, two intervalley backscattering channels are allowed at the potential steps, for pseudospin in $+x$ and $-x$ orientations respectively. At the Fermi energy ($E_f$), the propagation of pseudospin $|-x\rangle$ is through valence band states of wavevector $\pm(K - q_h)$ in the barrier region of width $W$, while in the well region of width $L - W$, it is through the conduction band states of wavevector $\pm(K + q_e)$.

Correspondingly, two scattering resonances occur for pseudospin $|-x\rangle$ in barrier region and well region respectively,

$$K - q_h = n\frac{\pi}{W}, \quad K + q_e = m\frac{\pi}{L-W}, \tag{1}$$

where $n$ and $m$ are integers. When both resonance conditions are satisfied under the constraint: $\hbar v_0(q_e + q_h) = U$, perfect transmission (Klein tunneling) is still expected for pseudospin $|-x\rangle$, as if the intervalley backscattering is absent. Similarly, pseudospin $|+x\rangle$ has scattering resonances,

$$K + q_h = n\frac{\pi}{W}, \quad K - q_e = m\frac{\pi}{L-W}, \tag{2}$$

The quantitative difference between Eq. (1) and (2) implies pseudospin dependent Klein tunneling that is controlled by $L$, $W$, and $U$.

The effect of intervalley backscattering is most significant under the maximum destructive interference in transmission (c.f. Fig. 1b). For pseudospin $|-x\rangle$, this occurs under

$$K - q_h = \left(n + \frac{1}{2}\right)\frac{\pi}{W}, \quad K + q_e = \left(m + \frac{1}{2}\right)\frac{\pi}{L-W}. \tag{3}$$

For pseudospin $|+x\rangle$, the condition becomes

$$K + q_h = \left(n + \frac{1}{2}\right)\frac{\pi}{W}, \quad K - q_e = \left(m + \frac{1}{2}\right)\frac{\pi}{L-W}. \tag{4}$$

As the number of periods increases, these conditions of maximum intervalley backscattering will introduce pseudospin dependent transport gaps that alternately appear at different energies in the minibands.

For a quantitative characterization of the superlattice minibands with the intervalley backscattering effects, we have calculated the energy dispersion and pseudospin textures using the tight-binding Hamiltonian,

$$H = \sum_i \varepsilon_i c_i^\dagger c_i - t \sum_{\langle i,j \rangle}(c_i^\dagger c_j + h.c.). \tag{5}$$

$c_i^\dagger$ ($c_i$) is the creation (annihilation) operator at site $i$, $\varepsilon_i$ the on-site potential which describes the superlattice potential, and the second term is the nearest neighbor hopping with $t \approx 2.8\text{eV}$. Periodic boundary conditions are applied in the $y$ direction, along which the electron's momentum $k_y$ is a conserved quantity.

Figs. 1c shows an example of the superlattice energy bands, with $U = 0.5\text{eV}$, $L = 40a \approx 10\text{nm}$, $W = 20a \approx 5\text{nm}$, $a$ being the lattice constant of graphene. The basic features include the anisotropic renormalization of the Dirac cones and generation of new Dirac points due to the intravalley scattering, as discovered in Ref. [7,31]. With the minima of the superlattice potential specified as energy 0 (c.f. Fig. 1a), the original Dirac points are at energy $E_D^0 = U/2$. The group velocity along $x$ remains at the pristine value $v_0$, and is reduced along $y$. New Dirac points are created at energies $E_D^n = E_D^0 \pm \hbar v_0 nG$, $G \equiv \frac{\pi}{L}$. The locations of the new Dirac points can be intuitively found from the schematic zone folding scheme of the superlattice dispersion at $k_y = 0$, as Fig. 1d illustrates.

Besides these new Dirac points, the zone folding also leads to intersections of bands of common pseudospin orientation. As highlighted in the zone folding scheme in Fig. 1d, the intersection of bands of pseudospin $|-x\rangle$ occurs at energies $E_-^n$ alternatively at the boundary and center of the mini-zone, while bands of pseudospin $|+x\rangle$ intersect at completely different set of energies $E_+^n$. The gapping of these intersection points requires intervalley scattering, since the two branches that run into each other are from the two valleys.

Our calculation accounting the intervalley scattering finds sizable gaps opened at these band crossings at $E_\pm^n$ (indicated by the arrows in Fig. 1c). Fig. 1e is a zoom-in of the band dispersion near energy 0 with a cut at $k_y = 0$, which includes a pair of Dirac points, a gapped crossing point for pseudospin $|+x\rangle$ in zone center, and a gapped crossing point for $|-x\rangle$ on zone boundary. These correspond respectively to the new Dirac points at energy $E_D^{-1}$, and the crossing points at energies $E_+^{-1}$ and $E_-^{-2}$, highlighted by the grey and green circles in the zone folding scheme in Fig. 1d. Within the staggered gaps $\Delta_+$ and $\Delta_-$, the minibands are pseudospin polarized. In particular, all states with $k_y = 0$ are fully polarized either in $|-x\rangle$ or $|+x\rangle$ pseudospin state. Remarkably, intervalley backscattering makes possible energy windows for polarized pseudospin transport in the superlattice.

Fig. 1f plots the pseudospin gaps as functions of $U$, fixing $L = 40a$ and $W = 20a$. Although the superlattice period $L$ is much larger than the lattice constant $a$, intervalley backscattering by the potential steps introduces sizable gaps of O(10) meV. The magnitudes of the gaps are nonmonotonic with the increase $U$, but rather have oscillations which are out-of-phase between the two pseudospin gaps. Remarkably, the zeros of $\Delta_-$ and $\Delta_+$ are given by the scattering resonances in Eq. (1) and Eq. (2) respectively for the cancellation of intervalley backscattering in individual period of the superlattice. Likewise, the maximal intervalley backscattering conditions Eq. (3) and Eq. (4) predict the maxima values of the gaps $\Delta_-$ and $\Delta_+$ respectively, while the envelope of these maxima is linear function of $U$. Equations (1-4) serve as simple guidelines to design superlattice potential for engineering pseudospin gaps at arbitrary Fermi energy.

**Valley-selective transmission through superlattice barrier.** When such superlattice is used as tunneling barrier, its staggered pseudospin gaps, combined with the valley-contrasted pseudospin chirality in pristine graphene, leads to the valley-selective Klein-tunneling. We consider finite periods of graphene superlattice connected on the two sides to semi-infinite pristine graphene leads, as shown in Fig. 2a. With the pseudospin textures of the Dirac cones in graphene leads, the incident electron has its valley index locked with the sublattice-pseudospin, i.e. K (-K) valley electron has pseudospin $|-x\rangle$ ($|+x\rangle$), for normal incidence from left. Transmission from one of the valleys can then be selectively blocked, due to the absence of the corresponding pseudospin state inside the staggered gaps in the superlattice barrier, as Fig. 2a illustrates.

The valley-dependent scattering by the superlattice barrier is calculated with the tight-binding Hamiltonian in Eq. (5), using a recursive Green's function technique.[32] Fig. 2b shows the calculated valley-conserved and valley-flip transmission and reflection coefficients under normal incidence ($k_y = 0$), for a 30-period superlattice barrier with $U = 0.5$eV, $L = 40a$, and $W = 20a$. In the shaded energy window that corresponds to the pseudospin gap $\Delta_+$ shown in Fig. 1e, we find perfect valley-conserving transmission for incidence in valley K, and perfect valley-flip reflection for incidence in valley -K. The energy window $\Delta_-$ exhibits the same behavior with opposite valley polarity for the allowed/blocked transmission. In the neighborhood of $\Delta_+$ and $\Delta_-$, multiple intervalley scatterings by the step-shaped edges in the potential also give rise to closely spaced scattering resonances which will develop into minibands in infinite period superlattice. The overall transmission probability $T \equiv (t_{K,K} + t_{-K,-K} + t_{K,-K} + t_{-K,K})/2$ equals 50% exhibiting the full valley selectivity within the pseudospin gaps, and approaches 1 far outside where perfect Klein tunneling is restored in both valleys.

Fig. 2c plots $P_v \equiv \frac{1}{2T}(t_{K,K} + t_{-K,K} - t_{-K,-K} - t_{K,-K})$, the valley polarization of the normal incidence transmission, as a function of the incident energy $E_f$ and barrier height $U$, for the same barrier geometry as in Fig. 2b. In the plotted range, we found four energy windows of valley-selective Klein tunneling, corresponding respectively to the pseudospin gaps $\Delta_+$ at $E_+^0$ and $E_+^{-1}$, and $\Delta_-$ at $E_-^{-1}$ and $E_-^{-2}$ illustrated in the zone folding scheme in Fig. 1d. The transmission has nearly perfect valley polarization in these pseudospin gaps with alternating valley polarity. Outside the gaps, the valley polarization sharply drops to zero. This makes possible sharp control of the valley filtering functionality of the junction by electrostatic control.

**Angle-integrated valley polarization.** The valley filter here functions in the coherent Klein tunneling regime, requiring the entire width of the superlattice to be small compared to the electron mean free path that can reach over 10 micron in high quality grapheme. [3,4] This allows the use of a few tens of periods, with $L \sim O(10)$ nm limited by the lithography of electrodes, which are sufficient to achieve high efficiency valley filter as Fig. 2d has shown. The filtered valley current, on other hand, can be harvested into channels beyond the ballistic limit as long as the valley polarity does not cancel after the integration over incidence/outgoing angle. The angle averaged valley polarity is therefore a key figure of merit for practical applications.

Fig. 3b shows the valley polarization $P_v$ and transmission probability $T$ as functions of the incident/outgoing angle $\phi = \tan^{-1}(k_y/k_x)$, for a 20-period superlattice barrier with $U = 0.5$eV, $L = 40a$, and $W = 20a$. Remarkably, a large valley polarization is achieved with the same sign over a significant range of angles. At angle $|\phi| > 60^o$ where $P_v$ drops to zero, the transmission probability $T$ has already become negligible. This leads to a pronounced angle-averaged valley polarization ($\bar{P}_v$) exceeding 70%, with an angle-averaged transmission probability $\bar{T} \sim 50\%$.

The wide-angle valley-filtering and the large $\bar{P}_v$ arise from the fact that the pseudospin gap exists over a range of $k_y$ that is comparable or larger than the lead Fermi surface. For the superlattice barrier used in the calculation of Fig. 3b, we show its energy contour inside the gap $\Delta_+$, color coded with the pseudospin texture, in Fig. 3a. The shaded area denotes the $k_y$ range in which pseudospin state $|+x\rangle$ is absent. As long as the Fermi surface in the graphene lead does not exceed this range, the conservation of $k_y$ and pseudospin in the scattering leads to valley filtering over the entire angle range where transmission is significant. This has been tested for several sizes of the Fermi surface in the graphene lead, tuned through an electrostatic shift $\delta$ of the lead Dirac points (c.f. Fig. 3a inset). As shown in Fig. 3b-c, the angle-averaged valley polarization remains large until the lead Fermi surface gets significantly larger than the shaded $k_y$ range in Fig. 3a, where both pseudospin states become available at large oblique angle. For the given superlattice, large angle-integrated valley filtering effect is expected for $\delta \lesssim 0.04$eV (Fig. 3c).

Fig. 3d-f show a further example of superlattice barrier, where the larger period $L = 120a \approx 30$nm leads to a narrower $k_y$ range of pseudospin polarized spectrum. Large $\bar{P}_v$ is obtained for $\delta \lesssim -0.04$ eV, and remarkably, by just using a 5-period barrier, the angle-averaged valley polarization can already reach ~ 20% (c.f. Fig. 3f).

**Discussion.** For electrostatically defined superlattice, the potential steps cannot be made atomically sharp. The lateral length scale of the potential step is determined by the vertical distance between the local gate and graphene, which can be made as small as ~ nm using hBN as the gate dielectric. [33] In general, superlattice with smoother potential steps will have smaller Fourier component responsible for the intervalley scattering. A reduction of the pseudospin gap size is therefore expected. To quantify such effect on the valley filtering, we calculate transmission through superlattice barrier with potential step described by hyperbolic function, i.e. $\frac{U}{2}\left[\frac{\tanh 6\ /d}{\tanh 3}+1\right]$ in $x \in \left[-\frac{d}{2},\frac{d}{2}\right]$, with $d$ characterizing the length scale (c.f. Fig. 4a inset).

Fig. 4a plots the pseudospin gap $\Delta_+$ as a function of $d$, for a superlattice with $L = 40a$, $W = 20a$ and $U = 0.5\text{eV}$. The gap size drops by half at $d = 8a \approx 2\text{nm}$. While this results in a narrower window for valley-selective tunneling, the effect on the performance inside the gap is less significant. Fig. 4b plots the valley filtering effect through 30 periods of the smoothened barriers. The angle-averaged valley polarization $\bar{P}_v$ shows a remarkable robustness to the smoothening of potential steps, still reaching ~ 70% at $d = 8a$.

We also examined the effect of statistical fluctuations in gate potential and gate width. The fluctuation of the gate potential is simulated by changing the height of barrier $i$ from $U$ to $U + \delta U_i$, where $\delta U_i$ is randomly chosen from a uniform distribution in a range $\left[-\frac{\delta U}{2},\frac{\delta U}{2}\right]$, with $\delta U$ characterizing the fluctuation strength. Similarly, fluctuation of the gate width is accounted by adding a random amount to the width of each barrier, with the fluctuation strength $\delta W$. We find the valley filter performance is remarkably robust against the fluctuations in the gate potential strength. Valley polarization $\bar{P}_v$ has negligible drop for $\delta U$ up to 20% of the barrier height $U$ (c.f. Fig. 4c). On the other hand, $\bar{P}_v$ is more sensitive to fluctuation of the gate width (Fig. 4d), which requires the lithography error to be controlled in ~ nm scale.

We also note that when the superlattice period $L$ is integer number of $3a$, the pseudospin gaps $\Delta_+$ and $\Delta_-$ are both located in the vicinity of the new Dirac points ($E_D^n$), as shown in the inset of Fig. 3d. The small offset between $\Delta_+$ and $\Delta_-$ comes from the trigonal warping of the graphene Dirac cone. In such case, with $\Delta_+$ and $\Delta_-$ close by in energy, an electrostatic switch of valley filtering polarity at fixed Fermi energy can be realized with a smaller change of barrier height.

## Acknowledgment

X.A. acknowledges support by Hebei Funds for Distinguished Young Scientists (No. A201808076) and Hebei Hundred Excellent Innovative Talents (No. SLRC2017035). W.Y. acknowledges support by Seed Funding for Strategic Interdisciplinary Research Scheme of HKU and Collaborative Research Funds (C7036-17W) of RGC.

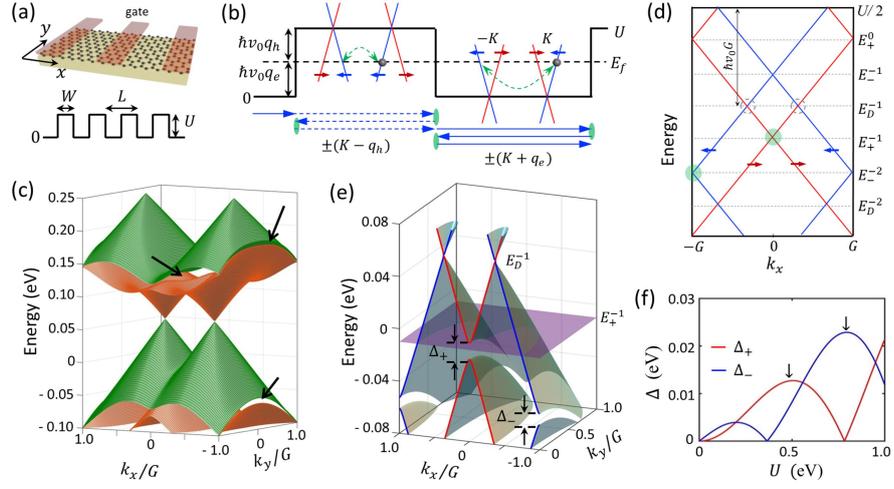

**Figure 1. Staggered pseudospin gaps by intervalley scattering**. (a) Schematic of an electrostatically defined superlattice potential. (b) Backscattering at the potential steps can only happen through the pseudospin-conserving intervalley process. Blue arrows denote the propagation in pseudospin $|-x\rangle$ state, with wavevector $\pm(K+q_e)$ and $\pm(K-q_h)$ respectively in the well and barrier regions. (c) An example of superlattice miniband dispersion, with $U = 0.5$ eV, $L = 40a \approx 10$ nm, $W = 20a \approx 5$ nm, $a$ being the lattice constant of graphene. (d) Schematic zone folding scheme, plotted at $k_y = 0$, for the superlattice in (c). States of $|+x\rangle$ and $|-x\rangle$ pseudospin are color coded in red and blue respectively. With energy 0 set at the potential bottom, the original Dirac points are at $U/2$. Intersections between bands of common pseudospin can be gapped when intervalley scattering is accounted. (e) Zoom-in of the minibands in (c) near energy 0. $\Delta_+$ and $\Delta_-$ are the gaps opened at the two intersection points highlighted by the green dots in (d). (f) $\Delta_+$ and $\Delta_-$ as functions of $U$, for the same superlattice periodicity.

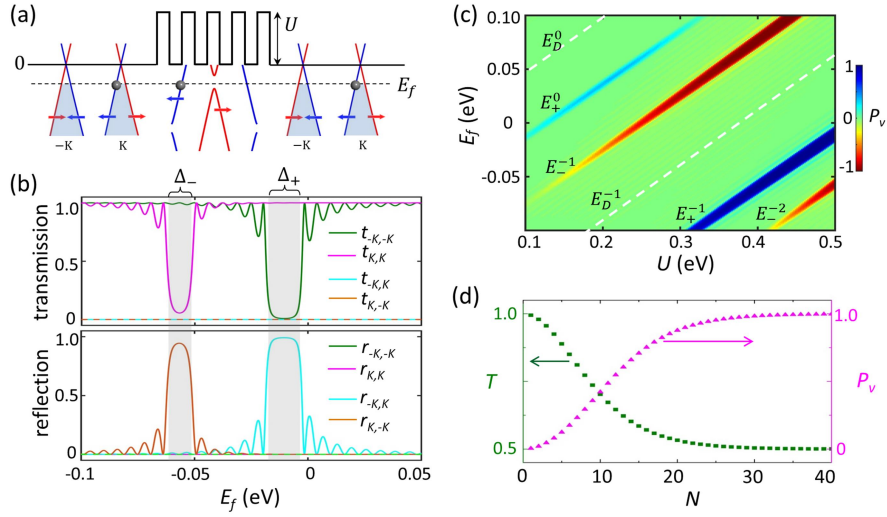

**Figure 2. Valley selective transmission through superlattice barrier.** **(a)** Schematic of a superlattice barrier connected to graphene leads on the two sides. For incidence at energy $E_f$ from left, transmission in the -K valley is blocked due to the absence of $|+x\rangle$ pseudospin states inside the $\Delta_+$ gap of the superlattice dispersion (c.f. Fig. 1e). **(b)** Valley-resolved transmission ($t$) and reflection ($r$) coefficients under normal incidence. Barrier height $U = 0.5$eV. **(c)** Valley polarization ($P_v$) of the transmission as a function of $U$ and Fermi energy $E_f$, under an unpolarized normal incidence. **(d)** Valley polarization and transmission probability $T$ as functions of superlattice period $N$. All calculations here use superlattice configuration $L = 40a \approx 10$nm, $W = 20a \approx 5$nm (c.f. Fig. 1a). $N = 30$ in (b) and (c).

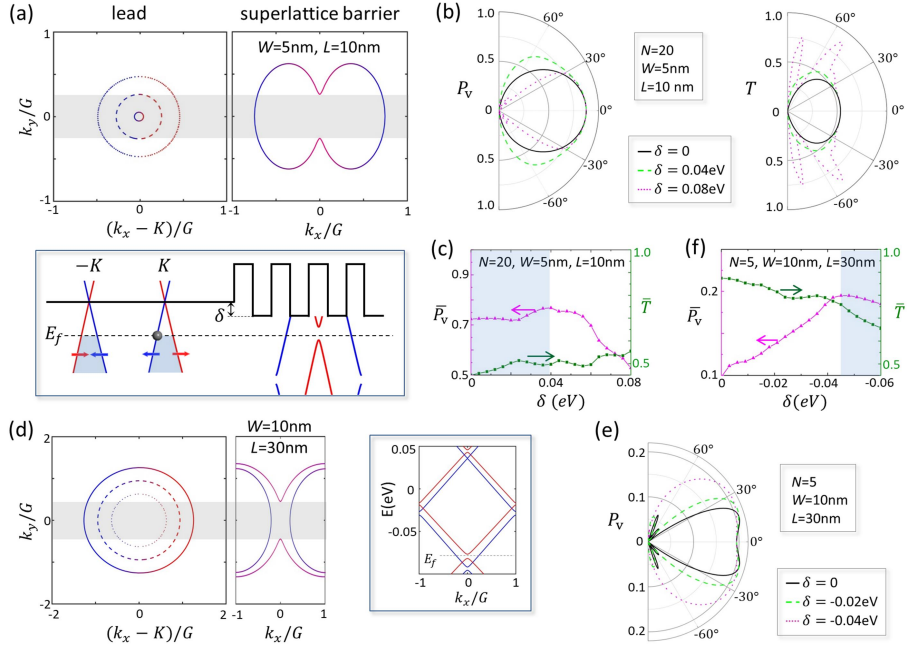

**Figure 3. Valley filtering performance under angle integration. (a)** Energy contour inside the $\Delta_+$ gap of the superlattice dispersion, in comparison with the lead Fermi surface which can be independently gate controlled. Solid, dashed and dotted lead Fermi surfaces correspond to parameter $\delta = 0$, 0.04. 0.08 eV respectively (c.f. inset). The pseudospin projections with $|+x\rangle$ and $|-x\rangle$ are color coded in red and blue respectively. **(b)** Valley polarization ($P_v$) and probability ($T$) of transmission as functions of incidence angle, at $\delta = 0$, 0.04, and 0.08 eV. **(c)** Angle-averaged valley polarization $\overline{P}_v$ and transmission probability $\overline{T}$, as functions of $\delta$. (b) and (c) use a 20-period superlattice barrier with $L = 40a \approx 10$nm, $W = 20a \approx 5$nm. **(d), (e), (f)** Similar plots for a superlattice of wider barriers and spacing: $L = 120a \approx 30$nm and $W = 40a \approx 10$nm, where the energy dispersion at $k_y = 0$ is shown in inset. (e) and (f) show the performance using only 5 periods of such superlattice barrier.

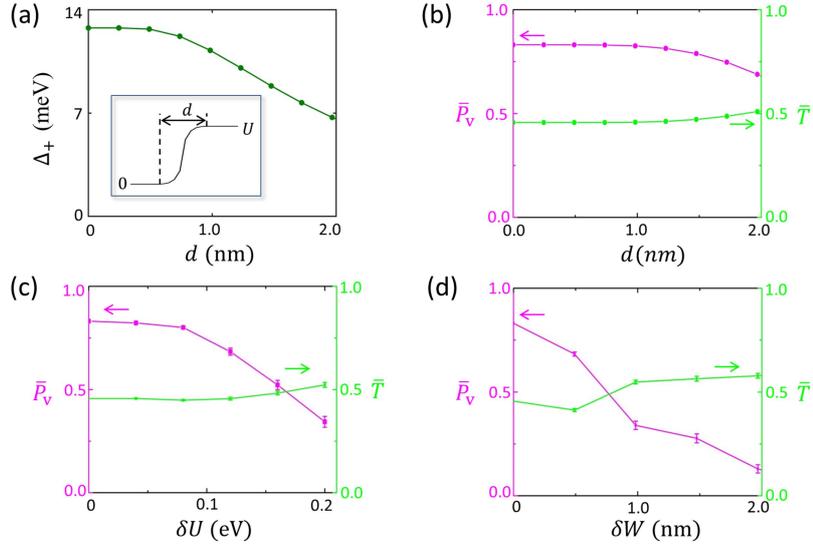

**Figure 4. Valley filtering performance under imperfections. (a)** Pseudospin gap $\Delta_+$ as a function of $d$, which quantifies the smoothening of the potential steps as shown in inset. **(b)** Angle-averaged valley polarization ($\bar{P}_v$) and probability ($\bar{T}$) of transmission, as functions of $d$. **(c)** $\bar{P}_v$ and $\bar{T}$ in presence of statistical fluctuation in barrier heights. **(d)** $\bar{P}_v$ and $\bar{T}$ in presence of statistical fluctuation in barrier widths. $L = 40a \approx 10$nm, $W = 20a \approx 5$nm, and 30 periods of such superlattice barrier is used in (b-d).